\documentclass[aps,prb,reprint,superscriptaddress]{revtex4-2}

\usepackage[utf8]{inputenc}
\usepackage[usenames, dvipsnames, svgnames, table]{xcolor}
\usepackage{mathtools}
\usepackage{amsfonts}
\usepackage{graphicx}
\usepackage[colorlinks=true, citecolor=Blue,
            linkcolor=BrickRed, urlcolor=ForestGreen]{hyperref}
\usepackage{siunitx}
\usepackage{xifthen}

\usepackage[shortlabels]{enumitem}

\DeclareSIUnit\ppm{ppm}

\newcommand{\refme}[1][]
{\textbf{\textcolor{red}{[r\ifthenelse{\isempty{#1}}{}{:#1}]\ }}}

\begin{document}


\title{Designing optimal linear detectors---a bottom-up approach}


\newcommand{\uob}{Institute for Gravitational Wave Astronomy, School of Physics and
Astronomy, University of Birmingham, Birmingham B15 2TT, United Kingdom}
\author{Joe Bentley}
\affiliation{\uob}
\email[Corresponding author: ]{joe.bentley@uni-hamburg.de}
\author{Hendra Nurdin}
\affiliation{School of Electrical Engineering and Telecommunications, University of New South Wales, Sydney 2052, Australia}
\author{Yanbei Chen}
\author{Xiang Li}
\affiliation{Theoretical Astrophysics 350-17, California Institute of Technology, Pasadena, California 91125, USA}
\author{Haixing Miao}
\affiliation{\uob}


\date{\today}

\begin{abstract}



This paper develops a systematic approach to realising linear detectors with an optimised sensitivity, allowing for the detection of extremely weak signals. First, general constraints are derived on a specific class of input-output transfer functions of a linear detector. Then a physical realization of transfer functions in that class is found using the quantum network synthesis technique, which allows for the inference of the physical setup directly from the input-output transfer function. By exploring a minimal realization which has the minimum number of internal modes, it is shown that the optimal such detectors are internal squeezing schemes. Then, investigating non-minimal realizations, which is motivated by the parity-time symmetric systems, a quantum non-demolition measurement is systematically recovered.
\end{abstract}


\maketitle

\section{Introduction}

The sensitivity of high-precision optical measurements, such as those performed in advanced gravitational wave (GW) detectors~\cite{AdvancedLIGO15,Acernese2015}, is ultimately limited by the fundamental fluctuations of the quantum vacuum, known as the quantum noise. For GW detectors, increasing the quantum-noise limited sensitivity will allow us to detect sources arising from a greater volume of the universe, as well the full neutron star inspiral waveform, allowing the determination of the neutron star equation of state~\cite{Read2009,Bauswein2012}. The sensitivity of linear detectors is ultimately constrained by the quantum Cramer-Rao Bound (QCRB) which states that the variance of the measured signal due to noise is inversely proportional the variance $\sigma_{NN}$ of the photon number of the probe degree of freedom coupled to the signal~\cite{Helstrom1967,Braginsky2000,Holevo2011,Tsang,Miaoa,Miao2019a}. For vacuum or coherent (laser) states of light, this quantity is ultimately limited by the Heisenberg limit $\sigma_{NN} \sim N^2$, which states that the uncertainty scales quadratically with the number of resources available (for optical detectors $N$ is the mean number of photons in the probe degree of freedom) although for most resonant detectors it is often constrained by the stronger shot-noise limit (also called the standard quantum limit in quantum metrology): $\sigma_{NN} = N$~\cite{Zwierz2012}. This shot-noise limit can be surpassed using techniques such as bandwidth broadening via negative dispersion~\cite{Pati2007,Yum2013,Miao2015,Bentley2019}, however in this case the Heisenberg limit is not saturated since $\sigma_{NN}$ is explicitly bounded to a value less than $N^2$. Previously the Heisenberg limit for phase measurement has been saturated in a non-resonant detector using a combination of entanglement, multiple sampling, and probabilistic adaptive measurements~\cite{Daryanoosh2018}. Theoretical examples of systems that saturate the limit have also been derived for exotic non-classical states~\cite{Tsarev2018, Huang2018} and using quantum error correction~\cite{Zhou2018a}. Here we instead focus on linear optical phase measurement, and focus on maximising $\sigma_{NN}$, however we cannot claim to saturate the Heisenberg limit due to our approximation of a linearized coupling of the signal to the detector. We call a linear detector with a maximal $\sigma_{NN}$ an \emph{optimal linear detector} since as we will see it also optimises the sensitivity to a linearly coupled classical signal. In this paper we then introduce a general approach to realising an optimal linear detector directly from its input-output transfer function.

A brief outline of the approach is as follows. First, we start with a physically realizable transfer function with order $n$ in frequency. This order sets a limit on the complexity of the system, since the corresponding physical realization will have a number of internal modes limited by this order. Then, from this realization, we investigate to which internal mode a signal should be coupled in order to gain a maximum signal-to-noise ratio, corresponding to the internal mode to which the input vacuum fluctuations have the maximum coupling, thus giving an optimal $n$-th order detector.

This approach leads to both minimal and non-minimal realizations: the minimal realization of a first-order transfer function that is active (has non-unity gain) exhibits internal squeezing, directly increasing the photon number fluctuation in the probe degree of freedom, as explored in~\cite{Peano2015, Korobko, Korobko2019}; the non-minimal realization begins with the minimal realization of a first-order lossless passive detector that is shot-noise limited, then adds a pair of auxiliary modes that result in an infinite signal amplification at DC (i.e.~for low frequency signals). The latter is motivated by the parity-time (PT) symmetric system explored in~\cite{Li2020,Li2021}. In both cases, we use the systematic realization framework developed in~\cite{Bentley2021} to realize the simplest single degree-of-freedom system with squeezing, which can be extended to arbitrarily many degrees-of-freedom using this framework. For the signal amplification case, the input-output relation must remain the same so that additional noise channels are not added, and we derive the corresponding conditions for modifying a system's internal dynamics without affecting its input-output transfer function. Additionally we show that this is related to an ideal QND (quantum non-demolition) measurement~\cite{Braginsky92, Yamamoto2014a}, and quantum-mechanics-free subsystems~\cite[\S III]{James2008}, \cite[Appendix D]{Gough2007}, \cite{Tsang2012, Wang2013}. Further we show how the dynamics can be arbitrarily modified by adding additional hidden modes while maintaining the QND property of the variable, for example so that the detector's most sensitive frequency can be tuned to a specific frequency, not limited to DC. As discussed, in all cases we start with the system's transfer function, since the order of a system's input-output transfer function determines the complexity of the minimal realization of the system, specifically determining the number of internal degrees of freedom of the system~\cite{Nurdin2009}. In this way we can then start with full control on the complexity of the resulting detector in the design process.

The outline of this paper is as follows. In Section.~\ref{sec:linear_sys_transf_func} we define the transfer function used to analyse the linear systems in this paper. In Section.~\ref{sec:qcrb} we will show how quantum network synthesis can be used to find a physical realization directly from the transfer function, and how the resulting detector's performance can be evaluated using the Quantum Cram\'{e}r-Rao bound. In Section.~\ref{sec:conditions} we will then discuss the various conditions on the transfer function arising from the physical realizability conditions, which specifies the number of parameters needed to describe a physically realizable system. In Section.~\ref{sec:internal-squeezing-general} we consider the optimal minimal realization of the first-order transfer function, showing that it is an internal squeezing scheme. In Section.~\ref{sec:qcrb:signal-amplification} we consider the optimal non-minimal by augmenting a shot-noise limited tuned cavity with auxiliary modes that do not affect the input-output dynamics, leading to a saturated signal amplification, and further that an ideal QND measurement is realized. It must be stressed that in both cases the infinite signal response arises from the approximations of our analysis, which breaks down as the uncertainty becomes comparable to the mean, and therefore we cannot infer that we completely reach the Heisenberg limit.

\section{The transfer function of linear systems}
\label{sec:linear_sys_transf_func}
In this work we are concerned with quantum systems with linear dynamics in the Heisenberg picture \cite{Nurdin2017}. In this section we briefly review some of the relevant concepts with further details deferred to Appendix \ref{appendix:lin-sys}. We consider finite-dimensional linear quantum systems (in the sense that there are only a finite number of internal modes) with  their dynamics given in the general form
\begin{align}
\begin{split}
\dot{\mathbf{x}} &= A \mathbf{x} + B\mathbf{u}\\
\mathbf{y} &= C \mathbf{x} + D\mathbf{u},
\end{split} \label{eq:lin-sys}
\end{align}
where $\mathbf{x}$ is a column vector containing the operators for the internal modes, $\mathbf{u}$ is a column vector of input operators to the system, and $\mathbf{y}$ is column vector of output operators from the system; $A$, $B$, $C$, and $D$ are complex matrices of the appropriate dimensions compatible with $\mathbf{u}$, $\mathbf{x}$ and $\mathbf{y}$. In particular, $A$ is a square matrix. More generally, $A$, $B$, $C$ and $D$ could potentially be integro-differential operators rather than constant matrices. In some cases this may be rewritten in the form \eqref{eq:lin-sys} by introducing a finite number of additional degrees of freedom. However, in general it may be necessary to introduce an infinite number of additional degrees of freedom, corresponding to $A$, $B$, $C$ and $D$ having an infinite number of rows and columns. An example is a model of linear gradient echo memories considered in \cite{HCHJ13}, a model involving continuous spatial degrees of freedom. For the purposes of this work, we are interested solely in the finite-dimensional case as given by \eqref{eq:lin-sys}.

In this paper we will restrict the analysis to single-input single-output (SISO) quantum systems with the input and output fields each described by a pair of bosonic annihilation and creation operators or quadratures in the two-photon formalism described by Caves and Schumaker~\cite{Caves1985, Schumaker1985}.  The output $\mathbf{y}$ has the decomposition $\mathbf{y}=\mathbf{y}_n+\mathbf{y}_f$, where $\mathbf{y}_n$ and $\mathbf{y}_f$ are the natural response and forced response, respectively,  given by:
\begin{eqnarray*}
\mathbf{y}_n(t) &=& Ce^{At} \mathbf{x}(0^-), \label{eq:natural}\\
\mathbf{y}_f(t) &=&\int_{0^-}^{t} Ce^{A(t-\tau)} B\mathbf{u}(\tau) d\tau + D\mathbf{u}(t),   \label{eq:forced}
\end{eqnarray*}
where $\mathbf{x}(0^-)$ is the initial condition for $x$ at time $t=0^-$ just before $t=0$. 
The natural response only depends on the initial condition $\mathbf{x}(0^-)$ but not the input $\mathbf{u}$ while the forced response is the system's output response to the input $\mathbf{u}$ and does not depend on $\mathbf{x}(0^-)$. 

For causal systems, the impulse response or Green's function $h$ for the system is given by  $h(t) = Ce^{At} B\, \Theta(t) + D\delta(t)$ where $\Theta(t)$ is the Heaviside step function and $\delta(t)$ is the Dirac delta function. Note that $\mathbf{y}_f$ is the convolution of $h$ with $\mathbf{u}$. If $h(t) = O(e^{\sigma t})$ then the system's transfer function is a complex function defined by: 
\begin{align}
H(\Omega) =\int_{0^-}^{\infty} h(t) e^{i\Omega t} \mathrm{d}t,\; \Im\{\Omega\}>\sigma. \label{eq:TF}
\end{align}
To conclude this section, we note that the transfer function coincides with the unilateral (one-sided) Laplace transform $\mathcal{L}[h](s) = \int_{0^-}^{\infty} h(t)e^{st} \mathrm{d}t$ of $h$  with the identification $s=i\Omega$ (note that in many fields, for example in engineering as in~\cite{Shaiju2012}\cite[Chapter 2]{Nurdin2017}, the Laplace transform is defined with $s$ replaced by $-s$). 

\section{Quantum Cram\'{e}r-Rao Bound}
\label{sec:qcrb}

In this section we show how an optimal detector can be engineered by minimising the Quantum Cram\'{e}r-Rao Bound (QCRB)~\cite{Miao}. This is performed by engineering diverging transfer functions from the vacuum input to the probe degree-of-freedom. As stressed previously, any divergent response we engineer is a result of our approximations of a linear signal coupling, as well as a more specific approximation arising from the specific physical realization of each system which we will treat as we come to them.

The QCRB sets a lower limit on the variance of an unbiased estimator of a classical signal $x_c(t)$ coupled to a detector linearly via $\hat{H}_\text{int} = -\hat{F}x_c(t)$,
\begin{equation}
	S_{xx}(\Omega) > \frac{\hbar^2}{ S_{FF}(\Omega)} \equiv \sigma^\text{QCRB}_{xx}(\Omega),\label{eq:qcrb:qcrb}
\end{equation}
where $S_{xx}(\Omega)$ is the single-sided displacement power spectral density and $S_{FF}(\Omega)$ is the spectral density describing the quantum fluctuations of the degree of freedom $\hat{F}$ that couples to the classical signal. In this paper we consider lossless systems, and so the spectral density of $\hat{F}$ is given by,
\begin{equation}
	S_{FF}(\Omega) = S_{u u}(\Omega) |G_{u F}(\Omega)|^2 = |G_{uF}(\Omega)|^2,\label{eq:qcrb:transfer-fn-to-ff}
\end{equation}
where $G_{u F}(\Omega)$ is the open-loop transfer function from the input $\hat{u}$ to the internal degree-of-freedom $\hat{F}$ which belongs to the vector of internal modes $\mathbf{x}$. With external squeezing we decrease the QCRB by increasing $S_{uu}(\Omega)$ which is a well known technique, so here we have restricted the input to a unsqueezed vacuum input: $S_{u u}(\Omega) = 1$.

We illustrate the general process using an optical interferometer with a Fabry-P\'{e}rot cavities which detects small modulations of the cavity lengths $x_c(t) = \Delta L(t)$, equivalent to differential displacement of the mirrors~\cite{Kimble2000}. In this case after linearizing the radiation pressure force of the light on the mirrors (as discussed in Section.~\ref{sec:internal-squeezing-general}) we arrive at a linear coupling of this length modulation to the amplitude quadrature of the cavity, so that the probe degree of freedom is $\hat{F} \propto \hat{a}^1 \equiv \hat{a} + \hat{a}^\dag$ where $\hat{a}$ is the annihilation operator of the cavity mode. Here the probe fluctuation is related to the intracavity photon number fluctuation by,
\begin{equation}
S_{FF}(\Omega) =|G_{uF}(\Omega)|^2= \frac{\hbar^2 \omega_0^2}{ L^2} S_{NN}(\Omega) ,
\end{equation}
where $S_{NN}(\Omega)$ is the spectral density describing the photon number fluctuations, $\omega_0$ is the laser carrier frequency, and $L$ is the arm cavity length. In this case we measure the displacement $x_c(t)$ and therefore a good figure of merit is the signal-to-noise ratio,
\begin{equation}
	\text{SNR} = \int_0^{\infty} \frac{\mathrm{d}\Omega}{2\pi} \frac{ |x_c(\Omega)|^2}{S_{xx}(\Omega)},
\end{equation}
where $x_c(\Omega)$ is the Fourier transform of the classical signal. For a displacement spectrum that is flat (frequency independent): $|x_c(\Omega)|=|x_c|$, this SNR is bound by the QCRB,
\begin{align}
    \text{SNR} \leq\ & \int_0^{\infty}\frac{\mathrm{d}\Omega}{2\pi} \frac{ |x_c|^2}{\sigma_{xx}^\text{QCRB}(\Omega)} = \label{eq:snr-qcrb} \\
    &\frac{|x_c|^2}{\hbar^2} \int_0^{\infty} \frac{\mathrm{d}\Omega}{2\pi}\, |G_{uF}(\Omega)|^2 
    = \frac{\omega_0^2 |x_c|^2}{L^2} \sigma_{NN},\nonumber
\end{align}
and $\sigma_{NN}$ is total variance of the photon number of the probe degree of freedom. Therefore by maximizing $|G_{uF}(\Omega)|^2$, we maximize the probe fluctuation $S_{FF}(\Omega)$ and the photon number variance $\sigma_{NN}$, therefore minimizing the QCRB and maximizing the SNR.

The general approach for realising an optimal detector is then performed as follows. First, as shown in~\cite{Bentley2021}, we can synthesise any $n$ degree-of-freedom system directly from its input-output transfer function, so long as it obeys certain conditions which will be discussed in Section.~\ref{sec:conditions}. Then, labelling each internal degree-of-freedom of the realization as $\hat{F}_i,\ i = 1,\dots,n$ we can then calculate the open-loop transfer functions from the input to those degrees of freedom, $G_{uF_i}(\Omega)$. Finally, we can then maximise the right-hand-side of Eq.~\eqref{eq:snr-qcrb} by maximising $G_{uF_i}(\Omega)$ for the optimal system parameters and also choosing the optimal internal degree-of-freedom $\hat{F}_i$ to couple the signal $x(t)$ to, giving us a systematic way of optimising the detector design given the input-output transfer function.


\section{General constraints on the input-output relations of linear detectors\label{sec:conditions}}

In this section we will derive some general constraints on a certain class of transfer functions by investigating their physical realizability. We do this by investigating whether or not they can lead to dynamics that preserve the commutation relations of the input-output operators~\cite{James2008,Shaiju2012}. 

In this work we consider a class of quadrature-picture rational transfer functions (i.e., with scalar rational transfer functions as entries) that can be transformed into diagonal form with rational entries through multiplication by unitary transfer functions on the left and on the right. However, in doing this we lose some generality, since not all transfer functions can be transformed into this form. In general, one can transform transfer functions into a diagonal form with non-rational entries (by applying the symplectic decomposition in \cite{LN03} to each value of $\Omega$). Nonetheless, non-rational transfer functions can be approximated arbitrary closely with rational transfer functions by using methods such as Pad\'{e} approximation \cite{Baker96} and requiring that the approximation satisfy the physical realizability condition \eqref{eq:j-j-unitary-quadrature} below. 


Let $\hat{y}^{1,2}$ and $\hat{u}^{1,2}$ be the output and input fields respectively in the two-photon quadrature formalism~\cite{Caves1985,Schumaker1985}. Then they are related by a transfer function $\mathbf{G}_q$ (a $\mathbb{C}^{2 \times 2 }$-valued complex function) as 
\begin{equation}
    \begin{bmatrix}
    \hat{y}_1(\Omega) \\
    \hat{y}_2(\Omega)
    \end{bmatrix}
    = \mathbf{G}_q(\Omega)
    \begin{bmatrix}
    \hat{u}_1(\Omega) \\
    \hat{u}_2(\Omega)
    \end{bmatrix},
\end{equation}
where, as mentioned, we assume $\mathbf{G}_q$ to be a diagonal matrix with rational elements,
\begin{equation}
 \mathbf{G}_q(\Omega) =
    \begin{bmatrix}
		G_{11}(\Omega) & 0 \\ 0 & G_{22}(\Omega)
	\end{bmatrix}
    \label{eq:qcrb:transfer-matrix-quadrature}
\end{equation}
One special case is where $|G_{11}(\Omega)|^2 = |G_{22}(\Omega)|^2 = 1$ in which case we have no squeezing (see Section.~\ref{sec:qcrb:signal-amplification}). To parameterise the transfer function, we consider an $n$ degree-of-freedom pole-zero form,
\begin{equation}
	G_{11}(\Omega) = \frac{\prod_{j=1}^{n} (i\Omega - z_j)}{\prod_{k=1}^{n} (i\Omega - p_k)},
\end{equation}
where $\{z_j\in\mathbb{C}\,|\,j=1,\dots,n\}$, $\{p_k\in\mathbb{C}\,|\,k=1,\dots,n\}$ are the zeros and poles respectively, without any element in common between the zeros and poles. The expansion to $n$ is chosen as the minimal state-space realization of an $n$-th order pole-zero transfer function in this case will have $n$ internal modes, setting a limit on the complexity of the resulting realization, however applying the realization framework may result in additional auxiliary modes~\cite{Bentley2021}. As shown in~\cite{Shaiju2012} the transfer function is physically realizable if,
\begin{equation}
    \mathbf{G}_q^\dagger(-\Omega^*) \Theta \mathbf{G}_q(-\Omega) = \Theta,
    \label{eq:j-j-unitary-quadrature}
\end{equation}
and,
\begin{equation}
    \Theta = \begin{bmatrix}0 & i \\ -i & 0\end{bmatrix},
\end{equation}
which up to a factor of $i$ is the simplest matrix that takes part in the symplectic condition. This condition restricts the conjugate transfer function to,
\begin{equation}
	G_{22}(\Omega) = \frac{\prod_{k=1}^{n} (-i\Omega - p_k^*)}{\prod_{j=1}^{n} (-i\Omega - z_j^*)},
\end{equation}
and so the poles/zeroes of $G_{22}$ are the conjugates of the zeroes/poles of $G_{11}$ respectively and the sign of the frequency is flipped. Since the real and imaginary parts of the poles and zeroes are independent we have in total $4 n$ independent parameters specifying our system.

The possible poles and zeroes can be further reduced by noting that the quadrature operators are real in the time-domain and so the transfer function must obey $\mathbf{G}_q(-\Omega) = \mathbf{G}^\dag_q(\Omega)$, which leads to the equation,
\begin{equation}
	\prod_{j,k=1}^n (-i\Omega - p_k) (i\Omega - z_j^*)
	= \prod_{j,k=1}^n (i\Omega - p_k^*) (-i\Omega - z_j),
\end{equation}
which can be expanded as,
\begin{equation}
	\sum_{j=1}^{2 n} a_j (i\Omega)^{j - 1}  = 0,
\end{equation}
where $a_j$ are algebraic combinations of the poles and zeroes. Therefore we have $a_j = 0,\ j = 1,\dots,2n$ and the number of independent parameters is reduced to $2n$.

\section{Internal Squeezing}
\label{sec:internal-squeezing-general}

\begin{figure}[htb]
	\centering
	\includegraphics[width=0.7\linewidth]{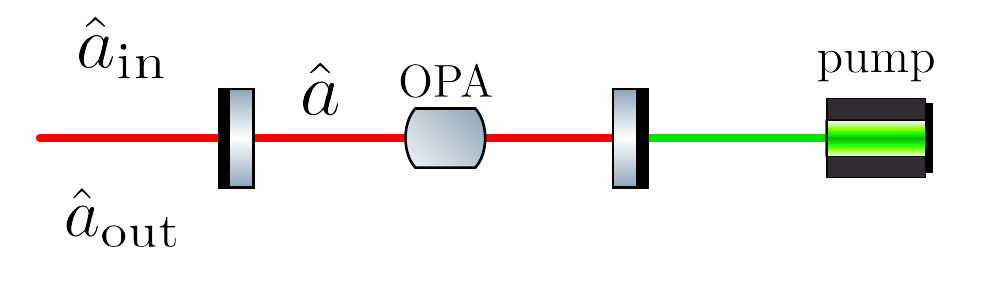}
	\caption{The setup for the squeezing of one quadrature via internal squeezing within the cavity, achieving an SNR that diverges at DC. The OPA (optical parametric amplifier) is pumped by the classical pump beam.\label{fig:internal-squeezing-setup}}
\end{figure}

In this section we consider the minimal realization of a first-order input-output transfer function exhibiting squeezing, i.e.~$|G(\Omega)| \neq 1$ for the quadrature operators, and the off-diagonal terms are non-zero for the sideband operators. We will see that the internal squeezing schemes are optimal for a classical signal coupled linearly to the cavity mode. We will choose the classical signal to be a displacement of the cavity length, as discussed in Section.~\ref{sec:qcrb}, although the analysis is general to linear detectors. According to the previous section, the most general first-order transfer function that is rational is quantified by two independent real parameters, and is given in the quadrature picture by,
\begin{equation}
	\mathbf{G}^q(\Omega) =
	\begin{bmatrix}
	\frac{\alpha +i \Omega }{\beta -i \Omega } & 0 \\
 0 & \frac{\beta +i \Omega }{\alpha -i \Omega }
	\end{bmatrix},
	\label{eq:qcrb:internal-squeezing-tf}
\end{equation}
which obeys Eq.~\eqref{eq:j-j-unitary-quadrature}. In Appendix.~\ref{appendix:internal-squeezing-physical-realisation-process} we derive the physically realizable state-space directly from the above transfer function giving,
\begin{align}
A&=
\frac{1}{2}
\begin{bmatrix}
 -\alpha -\beta & \alpha -\beta \\
 \alpha -\beta & -\alpha -\beta
\end{bmatrix},\\
B&= \sqrt{\alpha +\beta }\ I_{2\times2},\nonumber\\ 
C&= -\sqrt{\alpha +\beta }\ I_{2\times2},\
D= I_{2\times2}.\nonumber
\end{align}
The physical realization of this system has one internal degree of freedom and in the generalized open oscillator formalism (discussed extensively in~\cite{Nurdin2009}) is given by,
\begin{align}
    S &= I_{2\times2} \\
    \hat{L} &= \sqrt{\alpha+\beta}\,\hat{a} \\
    \hat{H} &= -\frac{i}{4}\hbar (\alpha-\beta)(\hat{a}\hat{a} - \hat{a}^\dag\hat{a}^\dag),
\end{align}
where $\hat{a}$ is the annihilation operator of the cavity mode. Here $S$ is the input-output direct scattering matrix, $\hat{L}$ is the coupling operator to the external continuum, and finally $\hat{H}$ is the system's internal Hamiltonian in the rotating frame at the laser carrier frequency. As shown in~\cite{Bentley2021} this corresponds to a tuned cavity with coupling coefficient $\gamma \equiv (\alpha+\beta)/2$ containing a non-linear crystal with coupling frequency $\chi \equiv (\alpha-\beta)/2$, which is related to the single-pass squeezing factor by $r = 2 \chi L / c$ where $L$ is the cavity length. Inverting these relations gives $\alpha = \gamma + \chi$ and $\beta = \gamma - \chi$.

The quadrature picture transfer function from the inputs to the internal degree of freedom is given by,
\begin{equation}
\begin{bmatrix}
	\hat{a}^1 \\
	\hat{a}^2
\end{bmatrix} = 
\begin{bmatrix}
 \frac{\sqrt{2 \gamma}}{\gamma - \chi -i \Omega } & 0 \\
 0 & \frac{\sqrt{2 \gamma}}{\gamma + \chi -i \Omega }
\end{bmatrix}
\begin{bmatrix} \hat{a}_\text{in}^1 \\ \hat{a}_\text{in}^2 \end{bmatrix}.
\end{equation}

In this case, the classical signal $x_c$, whose variance is lower-bounded by the QCRB in Eq.~\eqref{eq:qcrb:qcrb}, is the length modulation of the cavity, and is coupled to the probe degree of freedom linearly. This probe degree of freedom is, in turn, proportional to the probe amplitude quadrature $\hat{a}^1$. The linear coupling requires the application of a linearization procedure of the radiation pressure as performed in~\cite{Chen2013}, which specifically requires that any fluctuations of the field are much smaller than the mean. Fixing the proportional constant (the actual value does not affect the optimisation), we have $\hat{F} \equiv (\hbar\omega_0 \sqrt{2 N}/L)\hat{a}^1$ with $N$ the mean photon number and the input field is $\hat{u} \equiv \hat{a}_\text{in}^1$ giving the input-to-probe transfer function,
\begin{equation}
    G_{uF}(\Omega) =\left(\frac{\hbar\omega_0\sqrt{2N}}{L}\right) \frac{\sqrt{2 \gamma}}{\gamma - \chi -i \Omega }.
\end{equation}
The probe photon fluctuation is given by the integral in Eq.~\eqref{eq:snr-qcrb},
\begin{equation}
    \sigma_{NN} = 2N \int_0^{\infty} \frac{\mathrm{d}\Omega}{2\pi} \frac{2\gamma}{{|\gamma-\chi-i\Omega|}^2}.
\end{equation}
Clearly at $\gamma = \chi$ the integrand diverges at $\Omega = 0$ and thus the total probe fluctuation will diverge, corresponding to an increase in the signal response. As stressed previously, the probe fluctuation is only valid up to the point where it approaches the mean. This corresponds to the threshold case where the cavity acts as an optical parametric oscillator, since the damping of the cavity mode exactly compensates the pumping due to the non-linear crystal interaction. We have made the approximation that the pump will never be depleted, in practise as $\gamma$ approaches $\chi$, the latter will always change due to the decrease in pump power. At $\gamma \neq \chi$ the integral is solved trivially as,
\begin{equation}
    \sigma_{NN} =  \frac{N\gamma}{|\gamma - \chi|},
\end{equation}
which is shot-noise limited at $\chi = 0$, i.e.~at no internal squeezing, but can be made to surpass it for $\chi > 0$. We have therefore recovered the internal squeezing approach to enhancing the quantum-limited sensitivity previously developed in~\cite{Peano2015, Korobko, Korobko2019}.

In Appendix.~\ref{sec:quantum-expander} we demonstrate this approach starting with a second-order transfer function, arriving again at an internal squeezing design, with the same condition $\gamma = \chi$ resulting in a divergent response.

\section{Realization of QND via Signal amplification}
\label{sec:qcrb:signal-amplification}

\begin{figure}[htb]
    \centering
    \includegraphics[width=0.5\linewidth]{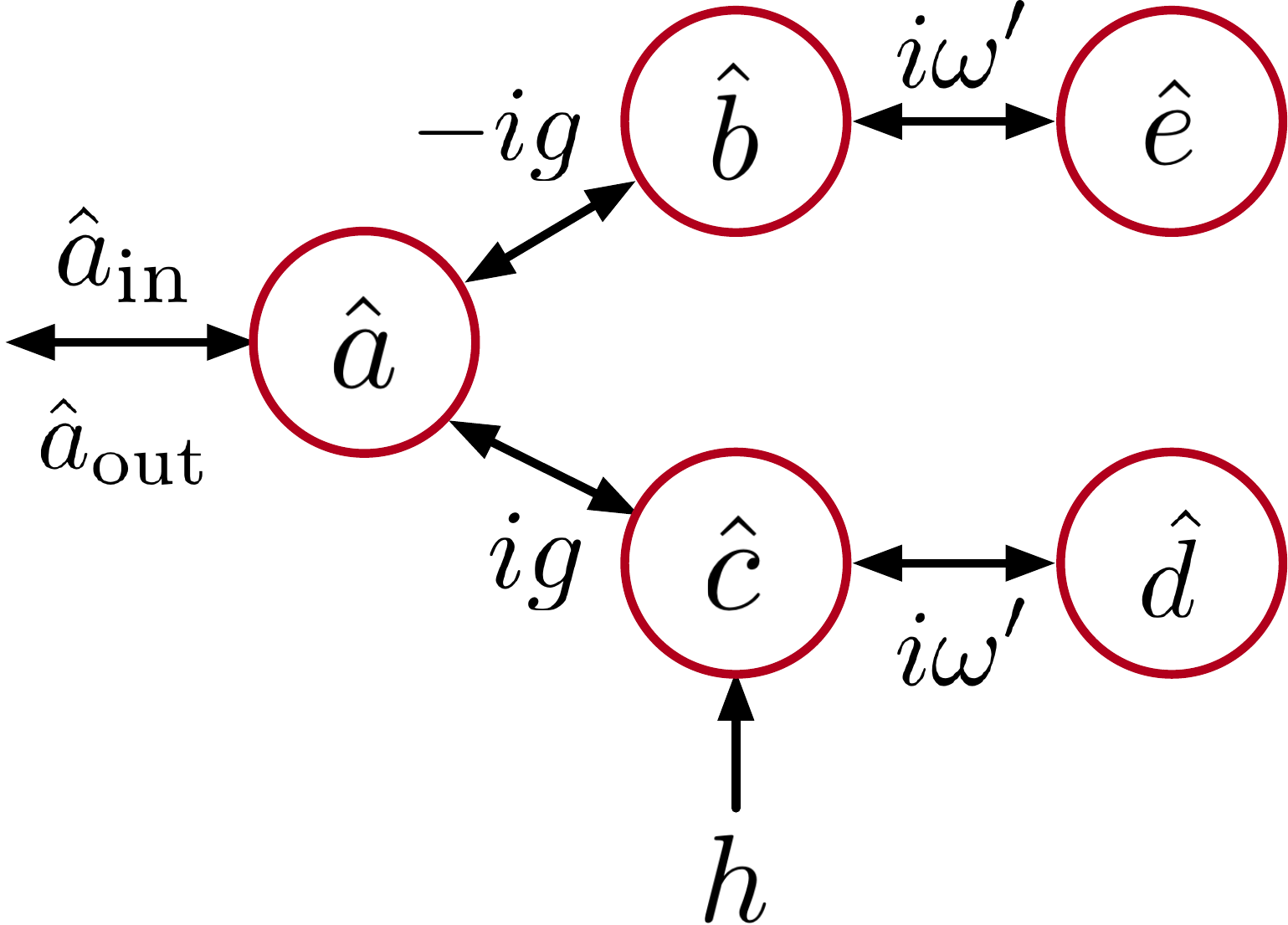}
    \caption{A schematic representation of the most complex realization developed in Section.~\ref{sec:qcrb:signal-amplification}. The modes $\hat{b}$ and $\hat{c}$ are coupled to the mode $\hat{a}$ via a squeezing-like interaction and beamsplitter-like interaction respectively with the same coupling frequency $g$, with the signal $h$ being coupled to mode $\hat{c}$. In this setup the signal response diverges at DC. We then add two additional modes $\hat{d}$ and $\hat{e}$ coupled to the modes $\hat{b}$ and $\hat{c}$ respectively, both via beamsplitter-like couplings with coupling frequency $\omega'$ which shifts the signal response resonance to $\Omega = \omega'$.}
    \label{fig:qcrb:signal-amplification}
\end{figure}

In this section we will consider the non-minimal realization of a passive first-order transfer function, showing that the shot-noise limit can be surpassed for a classical signal coupled linearly to one of the modes via the addition of hidden internal modes that internally amplify the signal. We will again choose the classical signal to be a displacement of the cavity length, as discussed in Section.~\ref{sec:qcrb}, although the analysis remains general to all linear detectors. We will first show that the minimal realization of a first-order system without squeezing is constrained by the Mizuno limit~\cite{JunMizuno}, and use the recent discovery that an infinite DC signal response can be achieved by adding a pair of modes that do not manifest in the input-output dynamics~\cite{Li2020} and thus do not add any additional noise. We will show how this is the simplest case of a general class of such non-minimal realizations, and that a quantum-mechanics-free subspace is formed allowing for arbitrary modification of the system dynamics. By asserting that the input-output behaviour remain the same, we have internal signal amplification without adding additional noise channels.

First, considering the case of Eq.~\eqref{eq:qcrb:internal-squeezing-tf} without squeezing with $\alpha = \beta \equiv \gamma$, the physically realizable state-space is given by,
\begin{equation}
    \mathbf{G}^q(\Omega) =
    \begin{bmatrix}
    \frac{\Omega - i\gamma}{\Omega + i\gamma} & 0 \\
    0 & \frac{\Omega - i\gamma}{\Omega + i\gamma}
    \end{bmatrix}.
\end{equation}
The corresponding generalized open oscillator is given by,
\begin{equation}
    S = I_{2\times2},\ \hat{L}= -\sqrt{2 \gamma}\hat{a},\ \hat{H} = 0,
\end{equation}
and the corresponding minimal realization is a simple tuned cavity where $\gamma$ is the cavity bandwidth. We can express the dynamics in the sideband picture by the Langevin equation and associated input-output relation,~\cite{Gardiner2004,Clerk2010,Chen2013,Aspelmeyer2014}
\begin{align}
    \dot{\hat{a}} &= -\gamma \hat{a} + \sqrt{2 \gamma} \hat{a}_\text{in}
    \label{eq:qcrb:langevin} \\
    \hat{a}_\text{out} &= \hat{a}_\text{in} - \sqrt{2 \gamma} \hat{a}.
\end{align}
As before, we choose the classical signal $x_c$ to be the length modulation of the $\hat{a}$ cavity, which is coupled linearly to the amplitude quadrature of the cavity, and therefore apply the aforementioned linearization process to the radiation pressure coupling. The transfer function from the input vacuum to the amplitude quadrature is given by,
\begin{equation}
    G_{uF}(\Omega) = \frac{\sqrt{2\gamma}}{\gamma - i\Omega}.
\end{equation}
Integrating this over all frequencies gives $2 \pi$ and thus the total SNR will be bounded by a constant independent of the bandwidth. This fact is the aforementioned Mizuno limit. Specifically, the detector is limited by the shot-noise limit $\sigma_{NN} = N$ where $N$ is the average photon number, giving $\text{SNR} = \omega_0^2 N^2 / 2 = E^2 / (2 \hbar^2)$ where $E$ is the average energy. Generally passive resonant detectors exhibit such a limit: dependent on the average stored energy, but independent of the parameters.

We now consider adding two auxiliary modes $\hat{b}$ and $\hat{c}$ as shown in Fig.~\ref{fig:qcrb:signal-amplification} and discussed in~\cite{Li2020}, without modifying the input-output dynamics. In Appendix.~\ref{appendix:qcrb:aux-mode-dynamics} we derive a general approach by which such hidden auxiliary modes can be added. The general condition is given by,
\begin{align}
    T^{(a)}(\Omega) \equiv &\begin{bmatrix} i g_b & -i g_{b^\dag} \\ i g_{b^\dag}^* & -i g_b^* \end{bmatrix} T^{(b)}(\Omega)\nonumber\\ + &\begin{bmatrix} i g_c & -i g_{c^\dag} \\ i g_{c^\dag}^* & -i g_c^* \end{bmatrix} T^{(c)}(\Omega) = 0.
    \label{eq:qcrb:full-transfer-matrix-main-text}
\end{align}
where $T^{(b)}(\Omega)$ and $T^{(c)}(\Omega)$ are transfer functions given in Eq.~\eqref{eq:qcrb:closed-transfer-b} and~\eqref{eq:qcrb:closed-transfer-c} respectively, and the most general linear Hamiltonian with two extra possibly-hidden modes is given by,
\begin{align*}
    &- \hbar g_b (\hat{a} \hat{b}^\dag + \hat{a}^\dag \hat{b})
    - \hbar g_{b^\dag} (\hat{a} \hat{b} + \hat{a}^\dag \hat{b}^\dag)\\
    &- \hbar g_c (\hat{a} \hat{c}^\dag + \hat{a}^\dag \hat{c})
    - \hbar g_{c^\dag} (\hat{a} \hat{c} + \hat{a}^\dag \hat{c}^\dag).
\end{align*}
The terms with coupling rates $g_b$ and $g_c$ are coupled to $\hat{a}$ via a beamsplitter, and the terms with coupling rates $g_{b^\dag}$ and $g_{c^\dag}$ via a non-linear crystal (or equivalently an optomechanical interaction with optomechanical coupling frequency $g$, as discussed in~\cite{Bentley2019, Li2020}). Each auxiliary mode has just one degree of freedom, and thus the transfer functions are given by,
\begin{align}
    T^{(b)}(\Omega) &= \frac{1}{-i\Omega} \begin{bmatrix}i g_b & -i g_{b^\dag} \\ i g_{b^\dag}^* & -i g_b^* \end{bmatrix}, \\
    T^{(c)}(\Omega) &= \frac{1}{-i\Omega} \begin{bmatrix}i g_c & -i g_{c^\dag} \\ i g_{c^\dag}^* & -i g_c^* \end{bmatrix}.
\end{align}
This gives,
\begin{equation}
    T^{(a)}(\Omega) \equiv \frac{1}{-i\Omega} (-g_b^2 + g_{b^\dag}^2 - g_c^2 + g_{c^\dag}^2) I_{2\times2}.
\end{equation}
If we now choose cavity mode $\hat{c}$ to be coupled to mode $\hat{a}$ purely by a beamsplitter-like interaction, then we have $g_c = \omega_s$ with $\omega_s$ being the sloshing frequency between the mode $\hat{c}$ and mode $\hat{a}$ and no non-linear coupling: $g_{c^\dag} = 0$. Therefore the input-output dynamics are left invariant if $g_{b} = 0$ and $g_{b^\dag} = \omega_s$, and so mode $\hat{b}$ should be coupled to mode $\hat{a}$ via a non-linear interaction (e.g.~a non-linear crystal if $\hat{b}$ is an optical mode) with the same coupling constant as the $\hat{c}$ mode: $g \equiv g_{b^\dag} = \omega_s$. The most general Hamiltonian with two hidden modes is thus given by,
\begin{equation}
    \hat{H}_0 = -\hbar g(\hat{a}\hat{c}^\dag+\hat{a}^\dag\hat{c}) - \hbar g(\hat{a}\hat{b} + \hat{a}^\dag \hat{b}^\dag).
    \label{eq:qcrb:signal-amplification-hamiltonian}
\end{equation}
Such a system is known as being PT(parity-time)-symmetric as the Hamiltonian is invariant under the parity operation (reversing modes $\hat{c}$ and $\hat{b}$ via $\hat{c} \leftrightarrow \hat{b}^\dag$, $\hat{c}^\dag \leftrightarrow \hat{b}$) together with the time reversal operation ($\hat{c} \leftrightarrow \hat{c}^\dag$, $\hat{b} \leftrightarrow \hat{b}^\dag$)~\cite{Bender2005}. The equations of motion are given by,
\begin{align}
    \dot{\hat{a}} &= -\gamma\hat{a} + ig \hat{c} + ig\hat{b}^\dag + \sqrt{2\gamma}\hat{a}_\text{in}, \\
    \dot{\hat{b}}^\dag &= -ig \hat{a}, \\
    \dot{\hat{c}} &= ig \hat{a}.
\end{align}
In this case, we have the probe degree of freedom proportional to the amplitude quadrature of the $\hat{c}$ mode cavity $\hat{F} \propto (\hat{c} + \hat{c}^\dag) / \sqrt{2}$ and the input as the amplitude quadrature $\hat{u} = (\hat{a}_\text{in} + \hat{a}_\text{in}^\dag) / \sqrt{2}$. Solving in the frequency domain we can find the input-to-probe transfer function,
\begin{equation}
    G_{uF}(\Omega) \propto \frac{g}{\Omega} \frac{\sqrt{2\gamma}}{\gamma - i\Omega},
\end{equation}
and therefore the probe fluctuation $S_{FF}$ diverges at DC and thus $\sigma_{NN}$ diverges and the detector is optimised in the regime where the photon number fluctuation is less than the mean. In this case we are not necessarily operating at or above threshold and thus the pump depletion approximation mentioned in the internal squeezing case is not relevant here.


We can show that the diverging probe fluctuation occurs due to the setup realising an ideal QND (quantum non-demolition) measurement~\cite{Braginsky92}, in which case the probe degree of freedom has infinite fluctuation as it is conjugate to a conserved QND quantity. Re-writing Eq.~\eqref{eq:qcrb:signal-amplification-hamiltonian},
\begin{equation}
    \hat{H}_0 = -\hbar g [\hat{a}^\dag (\hat{c}+\hat{b}^\dag) + \hat{a} (\hat{c}^\dag+\hat{b})],
\end{equation}
it can be seen that the composite quantity $\hat{c} + \hat{b}^\dag$ is conserved, since 
\begin{equation}
\frac{\mathrm{d}}{\mathrm{d}t}(\hat{c} + \hat{b}^\dag)\propto [\hat{c} + \hat{b}^\dag, \hat{c}^\dag + \hat{b}] = 0.
\end{equation}
This further implies two conserved quantities $\hat{X}_+ \equiv (\hat{X}_c + \hat{X}_b)/\sqrt{2}$, $\hat{Y}_- \equiv (\hat{Y}_c - \hat{Y}_b)/\sqrt{2}$ in terms of amplitude and phase quadratures $\hat{X}_c \equiv (\hat{c} + \hat{c}^\dagger) / \sqrt{2}$, $\hat{Y}_c \equiv (\hat{c} - \hat{c}^\dagger) / \sqrt{2}i$ (and similarly for $\hat{a}$ and $\hat{b}$). Additionally we also define $\hat{X}_- \equiv (\hat{X}_c - \hat{X}_b)/\sqrt{2}$. Rewriting the Hamiltonian in terms of these quantities gives,
\begin{equation}
    \hat{H}_0 = \hbar g(\hat{Y}_a \hat{X}_+ - \hat{X}_b \hat{Y}_-) - \hbar (\alpha /\sqrt{2}) (\hat{X}_+ + \hat{X}_-) h,
\end{equation}
The relevant residue part leading to detection of the signal $x_c$ reads:
\begin{equation}
    \hat{H}_\text{res} = -\hbar g\hat{X}_a \hat{Y}_- - \hbar (\alpha /\sqrt{2}) \hat{X}_- x_c,
\end{equation}
with $\hat{Y}_-$ being the conserved QND observable and $\hat{X}_-$ being the probe degree-of-freedom thus having infinite variance and therefore giving infinite signal response. In frequency-domain, $\hat{Y}_-$ exhibits the divergence at DC:
\begin{equation}
    \hat{Y}_-(\Omega) = \frac{i\alpha}{\sqrt{2}\Omega} x_c(\Omega).
    \label{eq:qcrb:yminus}
\end{equation}

As discovered above, the operators $\hat{X}_+$ and $\hat{Y}_-$ are constants of motion and therefore act form a quantum-mechanics-free subsystem~\cite[\S III]{James2008},\cite[Appendix D]{Gough2007},\cite{Tsang2012},\cite{Wang2013}. The dynamics can be arbitrarily modified while keeping this subsystem quantum-mechanics-free so long as the simultaneous measurability condition is kept,
\begin{equation}
    [\hat{X}_+(t), \hat{X}_+(t')] = [\hat{Y}_-(t), \hat{Y}_-(t')] = 0.
\end{equation}
We can then modify the dynamics such that the signal repsonse appears to diverge under the approximations of our analysis. As an example, we can shift the divergent response from DC to another frequency $\omega'$ by adding an extra pair of modes $\hat{d}$ and $\hat{e}$ which couple to $\hat{b}$ and $\hat{c}$ respectively. In this case the interaction Hamiltonian gains the following terms,
\begin{align*}
    -&\hbar \omega' (\hat{b}\hat{d}^\dag+\hat{b}^\dag\hat{d}+\hat{c}^\dag\hat{e}+\hat{c}\hat{e}^\dag) \\
    = &i\hbar \omega'(\hat{X}_+\hat{Q}_++\hat{Y}_+\hat{P}_+-\hat{Y}_-\hat{P}_-+\hat{X}_-\hat{Q}_-),
\end{align*}
which satisfies the general condition given in Eq.~\eqref{eq:qcrb:full-transfer-matrix-main-text} and thus does not affect the input-output dynamics, and where we have defined,
\begin{equation}
    \hat{Q}_\pm\equiv\frac{\hat{X}_d\pm\hat{X}_e}{\sqrt{2}},\quad
    \hat{P}_\pm\equiv\frac{\hat{Y}_d\pm\hat{Y}_e}{\sqrt{2}}.
\end{equation}
The residue part relevant to signal detection gains the term,
\begin{equation}
    \hbar\omega'(\hat{X}_-\hat{Q}_--\hat{Y}_-\hat{P}_-).
\end{equation}
The latter term modifies the dynamics of $\hat{Y}_-$ to become,
\begin{align}
    \dot{\hat{Y}}_- &= -\omega' \hat{Q}_- + \frac{\alpha}{\sqrt{2}} x_c,\\
    \dot{\hat{Q}}_- &= \omega' \hat{Y}_-,
\end{align}
and so eliminating $\hat{Q}_-$ in the frequency domain we obtain,
\begin{equation}
    \hat{Y}_-(\Omega) = \frac{i\alpha\Omega}{\sqrt{2}(\Omega^2-\omega'^2)} x_c(\Omega).
\end{equation}
We see that the signal response now diverges at $\Omega = \omega'$ rather than at DC, and that the PT symmetric case shown in Eq.~\eqref{eq:qcrb:yminus} is recovered for $\omega' = 0$. The final phase quadrature input-output relation is given by,
\begin{equation}
    \hat{Y}_\text{out}(\Omega) = -\frac{\Omega-i\gamma}{\Omega+i\gamma} \hat{Y}_\text{in}(\Omega) + \frac{\sqrt{\gamma}\alpha g\Omega x_c(\Omega)}{(\Omega^2-\omega'^2)(\Omega+i\gamma)},
\end{equation}
and so via the divergent signal amplification we now have an infinite signal response at a chosen frequency.

\section{Discussion}

We have demonstrated a systematic method for constructing detectors that are optimal up to the approximation that the photon number fluctuation approaches the mean, specifically investigating the first-order transfer function. Investigating minimal realizations of the transfer function, we showed that the optimal designs use internal squeezing. Considering non-minimal realizations, we investigated PT symmetric systems and showed that such systems realize a QND measurement and are optimal, further showing how their dynamics can be modified without losing this property.

Another further exploration will be a systematic realization of the transmission-readout setup presented in~\cite{Bentley2019}, however this requires systematically realising a $4\times4$ MIMO (multi-input multi-output) transfer function with third order elements, which will be complicated to realize directly using the aforementioned framework. Further, we can consider experimental demonstrations of the aforementioned PT-symmetric setups. Optomechanical realizations are currently in development, however there is also the open possibility for an all-optical demonstration that avoids the strict thermal noise requirements expected in the optomechanical design. Finally, the aforementioned signal amplification readout is less susceptible to output loss at the photodiode than the internal squeezing setup due to the amplification of the signal, so different realizations have different responses to loss and thus we can choose the specific realization to minimise the impact of loss on the SNR. One future approach would be a full analysis taking loss into account from the start by adding the additional loss channels to a MIMO transfer function, fixing the loss coefficient to be small, and then using the framework to systematically realize this transfer function.

\section{Acknowledgements}

We would like to thank Denis Martynov, LIGO AIC, and QNWG for fruitful discussions. 
J.B. is supported by STFC and School of Physics and
Astronomy at the University of Birmingham. J.B. and H.M.
acknowledge the additional support from the Birmingham
Institute for Gravitational Wave Astronomy.
H.M. has also been supported by UK STFC Ernest Rutherford 
Fellowship (Grant No. ST/M005844/11). Y.C. is supported by the Simons Foundation (Award Number 568762), and the National Science Foundation, through Grants PHY-1708212 and PHY-1708213. 

\appendix

\section{Linear Systems}  
\label{appendix:lin-sys}
The general solution for $\mathbf{x}$ and $\mathbf{y}$ of the linear system \eqref{eq:lin-sys} is:
\begin{eqnarray*}
\mathbf{x}(t) &=& e^{At} \mathbf{x}(0-) + \int_{0-}^{t} e^{A(t-\tau)} B\mathbf{u}(\tau) \mathrm{d}\tau,\\
\mathbf{y}(t) &=& Ce^{At} \mathbf{x}(0-) + \int_{0-}^{t} Ce^{A(t-\tau)} B\mathbf{u}(\tau) \mathrm{d}\tau + D\mathbf{u}(t).   
\end{eqnarray*}
Taking the initial condition to be at $t=0^-$ allows the consideration of impulse inputs $\delta(t)$ and a step discontinuity in $\mathbf{x}$ from $t=0^-$ to $t=0^{+}$ (the time just after $t=0$). 

The system is asymptotically stable if the matrix $A$ is Hurwitz, that is, all its eigenvalues have a real part $<0$. In an asymptotically stable system, the natural response asymptotically decays to zero , $\mathop{\lim}_{t \rightarrow \infty }e^{At}\mathbf{x}(0^-)=0$ for any initial condition $\mathbf{x}(0^-)$. The impulse response $h$ for an asymptotically stable system also decays to 0 as $t \rightarrow \infty$. 

For a sinusoid input $\mathbf{u}(t)= \mathbf{b}(\Omega)e^{-i\Omega t}$, where $\Omega$ is real and $\mathbf{b}(\Omega)$ is a fixed column vector of operators (in our case this vector contains the quantized modes of an input quantum field),  the asymptotic output  as $t \rightarrow \infty$ of an asymptotically stable system only has the forced response and is given by:
\begin{equation}
\mathbf{y}(t) = H(\Omega) \mathbf{b}(\Omega) e^{-i\Omega t}, \label{eq:forced-ss}
\end{equation}
where $H(\Omega)$ is as given in \eqref{eq:TF} with $\sigma \geq 0$. In this case, $H$ is coincides with the Fourier transform of $h$ and is called the system's frequency response. For an input $\mathbf{u}(t) = \int_{-\infty}^{\infty} \mathbf{b}(\Omega) e^{-i\Omega t}\mathrm{d}\Omega$, by linearity of the system the asymptotic response is $\mathbf{y}(t) = \int_{-\infty}^{\infty} H(\Omega) \mathbf{b}(\Omega) e^{-i\Omega t} \mathrm{d}\Omega$.

For a system that is not asymptotically stable, the impulse response  may not be integrable, $\int_{0^-}^{\infty} |h(t)|\mathrm{d}t = \infty$, and the frequency response $H$ not well-defined. For such systems, the forced response to a sinusoid input may not have an asymptotic solution. However, for $h(t)=O(e^{\sigma t})$, with $\sigma \geq 0$, it can have an asymptotic solution for inputs of the form $\mathbf{u}(t) = \mathbf{b}(\Omega)e^{-i\Omega t}$ for all complex $\Omega$ with $\Im\{\Omega\}>\sigma$ and with $\mathbf{b}(\Omega)$ some fixed vector of operators as before. In this case, the asymptotic forced response is again given  by the right hand side of  \eqref{eq:forced-ss} but $\Omega$ is now complex. For more general inputs of the form $\mathbf{u}(t) =\int_{-\infty}^{\infty} \mathbf{b}(\omega  + i\sigma_0) e^{-i(\omega  + i\sigma_0)t} \mathrm{d}\omega$, with $\sigma_0 > \sigma$, by linearity the forced response is given by $\mathbf{y}_f(t) = \int_{-\infty}^{\infty} H(\omega + i\sigma_0) \mathbf{b}(\omega + i\sigma_0) e^{-i(\omega + i\sigma_0)t} \mathrm{d}\omega$. 

\section{Physically Realizable State-Space Realization for Internal Squeezing\label{appendix:internal-squeezing-physical-realisation-process}}

We will now follow the steps given in~\cite{Bentley2021} to systematically construct the physically realizable state space of the internal squeezing setup whose transfer function is shown in Eq.~\ref{eq:qcrb:internal-squeezing-tf}. First we define a de-dimensionalised frequency with respect to $\alpha$ by making the transformation $\Omega \to \alpha \Omega$, so that the aforementioned transfer function becomes,
\begin{equation}
    \mathbf{G}^q(\Omega) =
	\begin{bmatrix}
	\frac{1 +i \Omega }{\Gamma -i \Omega } & 0 \\
 0 & \frac{\Gamma +i \Omega }{1 -i \Omega }
	\end{bmatrix},
\end{equation}
where $\Gamma \equiv \beta / \alpha$. The recovery of a linear quantum system \eqref{eq:lin-sys} with $\mathbf{G}^q(\Omega)$ as its transfer function is studied in modern control theory as a fundamental topic known as realization theory; see, e.g., \cite{deSchutter00} for a review. We do not describe the calculations for finding a state-space realization for the transfer function given above but we describe below how it may  can computed with Mathematica routines.

Assuming that the amplitude quadrature is squeezed, i.e.~$\Gamma < 1$, we can compute the canonical state-space realization using Mathematica. The Mathematica function \texttt{StateSpaceModel} when applied to a transfer function returns a state space in the controllable canonical form~\cite{wolfram_2021_statespacemodel,Luenberger1967}. The returned state space may have more internal degrees of freedom than is necessary to represent the system. This is remedied by applying the function \texttt{MinimalStateSpaceModel}~\cite{wolfram_2021_minimalstatespacemodel} to the state space to find the minimal state space~\cite{Antoniou1988}. Using these functions we calculate,
\begin{align}
    A'&=\frac{1}{3+\gamma^2} \begin{bmatrix}
        - (1+\Gamma)^2 & c_1 \\ c_1 & - 2 - \Gamma - \Gamma^3
    \end{bmatrix}, \\
    B'&= \begin{bmatrix}
        0 & \sqrt{\frac{2}{3+\Gamma^2}} \\ \frac{1}{2} \sqrt{\frac{3+\Gamma^2}{1+\Gamma^2}} & \frac{\Gamma^2 - 1}{2 \sqrt{3 + 4\Gamma^2 + \Gamma^4}}
    \end{bmatrix}, \\
    C'&= \begin{bmatrix}
        \frac{(1+\Gamma)^2 |\Gamma-1|}{\sqrt{2(3+\Gamma^2)}} &
        \frac{2(1+\Gamma)(1+\Gamma^2)}{\sqrt{3 + 4\Gamma^2 + \Gamma^4}} \\
        (1+\Gamma)\sqrt{\frac{3+\Gamma^2}{2}} & 0
    \end{bmatrix}, \\
    D'&= -I_{2\times2}.
\end{align}
where $c_1 = \sqrt{2} \sqrt{1+\Gamma^2} |\Gamma-1|$ and $I_{2\times2}$ is the $2\times2$ identity matrix. This state-space does not currently fulfil the physical realizability condition, given by, \cite{Bentley2021}
\begin{align}
    A J + J A^\dag + B J B^\dag &= 0, \label{eq:qcrb:physically-realisable-state-space-1} \\
    J C^\dag + B J D^\dag &= 0, \label{eq:qcrb:physically-realisable-state-space-2}
\end{align}
where,
\begin{equation}
    J = \begin{bmatrix}
        1 & 0 \\ 0 & -1
    \end{bmatrix}.
\end{equation}
To find the transformation from the unrealizable state-space $(A',B',C',D')$ to the realizable state-space $(A,B,C,D)$ we look for the matrix $X$ that satisfies,
\begin{align}
    A'X + X (A')^\dag + B' J (B')^\dag &= 0, \\
    X(C')^\dag + B' J (D')^\dag &= 0.
\end{align}
In this case one such matrix is given by,
\begin{equation}
    X = \begin{bmatrix}
        -\frac{2}{3+3\Gamma+\Gamma^2+\Gamma^3} & \frac{\sqrt{3+4\Gamma^2+\Gamma^4} |\Gamma-1|}{\sqrt{2}(1+\Gamma^2)(3+\Gamma^2)^{3/2}} \\
        \frac{1-\Gamma}{\sqrt{2(1+\Gamma^2)}(3+\Gamma^2)} & \frac{2}{3+3\Gamma+\Gamma^2+\Gamma^3}
    \end{bmatrix}.
\end{equation}
We then look for the similarity transformation matrix $T$ satisfying $X = T J T^\dag$, which in this case is given by,
\begin{equation}
    T = \begin{bmatrix}
        0 & \sqrt{\frac{2}{3+3\Gamma+\Gamma^2+\Gamma^3}} \\ \frac{1}{2} \sqrt{\frac{3+\Gamma^2}{1+\Gamma+\Gamma^2+\Gamma^3}} & \frac{\Gamma^2 - 1}{2 \sqrt{(1+\Gamma^2)(3+3\Gamma+\Gamma^2+\Gamma^3)}}
    \end{bmatrix}.
\end{equation}
We can apply this transformation by the standard state-space transformation,
\begin{equation}
    A = T^{-1} A' T,\ B=T^{-1}B',\ C=C'T,\ D=D',
\end{equation}
which gives the state-space,
\begin{align}
    A &= \frac{1}{2} \begin{bmatrix}
        -1-\Gamma & 1-\Gamma \\ |-1+\Gamma| & -1-\Gamma
    \end{bmatrix}, \\
    B &= \sqrt{1 + \Gamma} I_{2\times2}, \\
    C &= \sqrt{1 + \Gamma} I_{2\times2}, \\
    D &= -I_{2\times2}.
\end{align}
To slightly simplify the physical realization without loss of generality we can add a $\pi$ phase shift for the input-output reflection, resulting in transforming the state space via $C \to -C$ and $D \to -D$. Substituting $\Gamma = \beta / \alpha$ and reversing the de-dimensionalisation via $\Omega \to \Omega / \alpha$ we obtain the physically realizable state-space which obeys Eqs.~\eqref{eq:qcrb:physically-realisable-state-space-1} and~\eqref{eq:qcrb:physically-realisable-state-space-2},
\begin{align}
A&=
\frac{1}{2}
\begin{bmatrix}
 -\alpha -\beta & \alpha -\beta \\
 \alpha -\beta & -\alpha -\beta
\end{bmatrix},\\
B&= \sqrt{\alpha +\beta }\ I_{2\times2},\nonumber\\ 
C&= -\sqrt{\alpha +\beta }\ I_{2\times2},\
D= I_{2\times2}.\nonumber
\end{align}

\section{Quantum Expander}
\label{sec:quantum-expander}

\begin{figure}[h]
\centering
\includegraphics[width=0.6\linewidth]{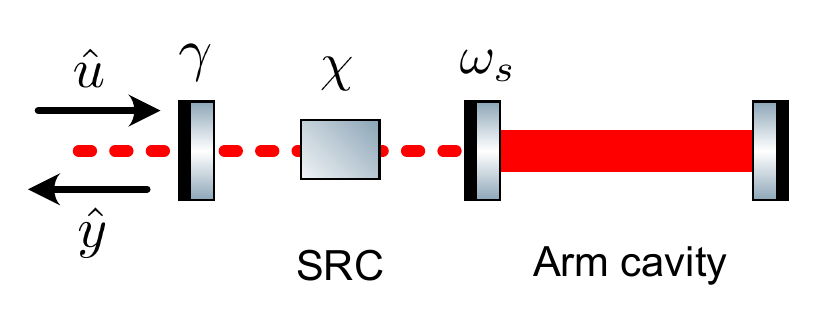}
\caption{The setup analysed for the quantum expander as explored in~\cite{Korobko2019}, equivalent to a tuned Michelson interferometer except with squeezing (via a non-linear crystal) internally within the SRC (signal recycling cavity).\label{fig:quantum-expander-setup}}
\end{figure}

In this appendix we consider a second-order input-output transfer function, showing that the optimal sensitivity is achieved when the parameters match that of the so-called quantum expander first explored in~\cite{Korobko2019}, a setup which, similarly to the transmission-readout setup discussed in~\cite{Bentley2019}, can directly increase the detection bandwidth of a gravitational wave interferometer. This setup, shown in Figure.~\ref{fig:quantum-expander-setup}, consists of a tuned, signal-recycled Michelson interferometer with internal squeezing in the signal recycling cavity. The signal-recycled Michelson can be mapped to an equivalent coupled-cavity~\cite{Buonanno2003}. We show that the quantum expander is the optimal detector for any second-order quadrature-picture transfer function obeying the constraints in Section.~\ref{sec:conditions}.

Thus we start with the most general rational and proper second-order input-output transfer function that is diagonal and rational,
\begin{equation}
\mathbf{G}^q(\Omega) =
	\begin{bmatrix}
	\frac{(i\Omega - \alpha_1) (i\Omega - \beta_1)}{(i\Omega - \alpha_2) (i\Omega - \beta_2)}
	& 0 \\
	0 & \frac{(-i\Omega - \alpha_2) (-i\Omega - \beta_2)}{(-i\Omega - \alpha_1) (-i\Omega - \beta_1)}
	\end{bmatrix},
\end{equation}
where $\alpha_1, \alpha_2, \beta_1, \beta_2 \in \mathbb{R}$. Requiring that there is no gain at DC we also obtain the condition $\alpha_1 \beta_1 = \alpha_2 \beta_2$. We then follow the same procedure as in Appendix.~\ref{appendix:internal-squeezing-physical-realisation-process} for finding the physically realizable state-space, obtaining,
\begin{align}
A&=
\begin{bmatrix}
 0 & 0 & - i \omega_s & 0 \\
 0 & 0 & 0 & i \omega_s \\
 - i \omega_s & 0 & -\gamma  & -\chi  \\
 0 & i \omega_s & -\chi  & -\gamma  \\
\end{bmatrix},\nonumber\\B&=
\begin{bmatrix}
 0 & 0 \\
 0 & 0 \\
 \sqrt{2} \sqrt{\gamma } & 0 \\
 0 & \sqrt{2} \sqrt{\gamma } \\
\end{bmatrix},\\
C&=
\begin{bmatrix}
 0 & 0 & -\sqrt{2} \sqrt{\gamma } & 0 \\
 0 & 0 & 0 & -\sqrt{2} \sqrt{\gamma } \\
\end{bmatrix},\ D=I_{2\times2}\nonumber,
\end{align}
where $\gamma \equiv \frac{1}{2}(-\alpha_1+\alpha_2-\beta_1+\beta_2)$, $\chi \equiv \frac{1}{2}(\alpha_1+\alpha_2\beta_1+\beta_2)$, and $\omega_s \equiv \sqrt{\alpha_1 \beta_1}$. This corresponds to the dynamics derived from Hamiltonian for the quantum expander first described in~\cite{Korobko2019}. In this case since we have a two degree-of-freedom we have to apply the separation theorem of~\cite{Nurdin2009} to separate it into two separated one degree-of-freedom systems. The corresponding quantum network is given by $\mathcal{N} = \{\{G_1, G_2\}, \hat{H}^d, \mathcal{S}\}$ where $\mathcal{S} = G_2 \triangleleft G_1$ represents the series product~\cite{Gough2007}, i.e.~the output of $G_1$ is fed into $G_2$. The two generalized open oscillators are given by,
\begin{align}
    G_1 &= \left(I_{2\times2}, 0, 0\right),\\
    G_2 &= \left(I_{2\times2}, -\sqrt{2\gamma} \hat{a}_q, \tfrac{i}{2} \hbar \chi (\hat{a}_q \hat{a}_q + \hat{a}_q^\dagger \hat{a}_q^\dagger) \right),
\end{align}
where $I_{2\times2}$ is the $2\times2$ identity matrix, $\gamma$ is the coupling frequency of the continuum to the cavity mode described by annihilation operator $\hat{a}_q$, and $\chi$ is the strength of the non-linear interaction. The Hamiltonian coupling the two cavities is given by $\hat{H}^d = \hbar \omega_s (\hat{a}_q \hat{a}^\dagger + \hat{a}_q^\dagger \hat{a})$ where $\hat{a}$ is the cavity mode of the second cavity, and is therefore a simple beamsplitter-like coupling between the two cavities. Note that $G_1$ is not coupled to the external continuum and therefore it is only coupled to $G_2$ via $H^d$. In total we have two tuned cavities coupled by a beamsplitter-like interaction, with the first cavity coupled to the external continuum and exhibiting internal squeezing, and have thus recovered the quantum expander realization pictured in Fig.~\ref{fig:quantum-expander-setup}.

The quadrature transfer function from the input to the arm cavity mode $\hat{a}$ was found to be,
\begin{equation}
\begin{bmatrix}
	\hat{a}^1 \\[0.5em]
	\hat{a}^2
\end{bmatrix} = 
\begin{bmatrix}
0 & \frac{\sqrt{2 \gamma } \omega _s}{i \omega  (\chi -\gamma )+\omega
   _s^2-\omega ^2} \\
 -\frac{\sqrt{2 \gamma } \omega _s}{-i \omega  (\gamma +\chi )+\omega _s^2-\omega
   ^2} & 0
\end{bmatrix}
\begin{bmatrix} \hat{a}_\text{in}^1 \\[0.5em] \hat{a}_\text{in}^2 \end{bmatrix}.
\end{equation}
Using Eq.~\eqref{eq:snr-qcrb} we see that the SNR for a signal coupled to the amplitude quadrature is given by $2 \pi \gamma / |\gamma - \chi|$, which diverges as $\chi \to \gamma$. At $\chi \gg \gamma$ the SNR approaches zero since the non-linear interaction totally depletes the amplitude quadrature fluctuations in the cavity. The SNR for the phase quadrature is given by $2 \pi \gamma / (\gamma + \chi)$ which is maximal at $\chi = 0$ where it is equal to $2\pi$ and is thus constrained by the Mizuno limit.

For the signal recycling cavity mode $\hat{a}_q$, we also see the divergence at $\chi \to \gamma$, except that in this case the SNR for the phase quadrature diverges rather than the amplitude quadrature,
\begin{equation}
\begin{bmatrix}
	\hat{a}_q^1 \\[0.5em]
	\hat{a}_q^2
\end{bmatrix} = 
\begin{bmatrix}
 -\frac{i \sqrt{2} \sqrt{\gamma } \omega }{-i \omega  (\gamma +\chi )+\omega _s^2-\omega
   ^2} & 0 \\
 0 & -\frac{i \sqrt{2} \sqrt{\gamma } \omega }{i \omega  (\chi -\gamma )+\omega
   _s^2-\omega ^2}
\end{bmatrix}
\begin{bmatrix} \hat{a}_\text{in}^1 \\[0.5em] \hat{a}_\text{in}^2 \end{bmatrix}.
\end{equation}
In this case the SNR for the amplitude quadrature is given by $2 \pi \gamma / (\gamma + \chi)$ and for the phase quadrature is given by $2 \pi \gamma / |\gamma - \chi|$ i.e.~the role of the amplitude and phase quadrature are swapped compared to the arm cavity mode.

\section{Auxiliary Mode Dynamics\label{appendix:qcrb:aux-mode-dynamics}.}

In this section we will discuss how the dynamics of added auxiliary modes, shown in Fig.~\ref{fig:qcrb:signal-amplification}, can be inferred by requiring that the frequency-domain input-output relation remain unchanged by the addition of them. Each auxiliary mode is coupled to a set of $n_d$ internal modes ${\hat{d}_j}$ and $n_e$ internal modes ${\hat{e}_j}$ adding the following terms to the Hamiltonian,
\begin{align*}
    \sum_j &- \hbar g_{d_j} (\hat{b} \hat{d}_j^\dag + \hat{b}^\dag \hat{d}_j)
    - \hbar g_{d_j^\dag} (\hat{b} \hat{d}_j + \hat{b}^\dag \hat{d}_j^\dag) \\
    &- \hbar g_{e_j} (\hat{c} \hat{e}_j^\dag + \hat{c}^\dag \hat{e}_j)
    - \hbar g_{e_j^\dag} (\hat{c} \hat{e}_j + \hat{c}^\dag \hat{e}_j^\dag)\\
    &+ \sum_{i\neq j} \hbar g_{d_i d_j^\dag} (\hat{d}_i \hat{d}_j^\dag + \hat{d}_i^\dag \hat{d}_j)
    + \hbar g_{d_i d_j} (\hat{d}_i \hat{d}_j + \hat{d}_i^\dag \hat{d}_j^\dag)\\
    &+ \sum_{i\neq j} \hbar g_{e_i e_j^\dag} (\hat{e}_i \hat{e}_j^\dag + \hat{e}_i^\dag \hat{e}_j)
    + \hbar g_{e_i e_j} (\hat{e}_i \hat{e}_j + \hat{e}_i^\dag \hat{e}_j^\dag),
\end{align*}
where $g_{d_i d_j^\dag}$ and $g_{d_i d_j}$ respectively quantify the beamsplitter-like and non-linear coupling between modes $\hat{d}_i$ and $\hat{d}_j$, and similarly for the $\hat{e}$ modes. Note that there is no direct coupling between the $\hat{d}$ and $\hat{e}$ modes.

The full set of equations of motion are,
\begin{align}
\dot{\hat{a}} = &-\gamma \hat{a} + \sqrt{2\gamma} \hat{a}_\text{in} \nonumber \\ &-i g_b\hat{b} + i g_{b^\dag} \hat{b}^\dag -i g_c\hat{c} + i g_{c^\dag} \hat{c}^\dag, \\
    \dot{\hat{b}} &= ig_b\hat{a} - ig_{b^\dag}\hat{a}^\dag + i \sum_j g_{d_j} \hat{d}_j - i \sum_j g_{d_j^\dag} \hat{d}_j^\dag, \label{eq:qcrb:b-dot} \\
    \dot{\hat{d}}_j &= ig_{d_j} \hat{b} + ig_{d_j^\dag} \hat{b}^\dag - i \sum_{i \neq j} g_{d_i d_j^\dag} \hat{d}_i
     - i \sum_{i \neq j} g_{d_i d_j} \hat{d}_i^\dag, \label{eq:qcrb:d-dot} \\
    \dot{\hat{c}} &= ig_c\hat{a} - ig_{c^\dag}\hat{a}^\dag + i \sum_j g_{e_j} \hat{e}_j - i \sum_j g_{e_j^\dag} \hat{e}_j^\dag, \\
    \dot{\hat{e}}_j &= ig_{e_j} \hat{c} + ig_{e_j^\dag}  \hat{c}^\dag - i \sum_{i \neq j} g_{e_i e_j^\dag} \hat{e}_i
     - i \sum_{i \neq j} g_{e_i e_j} \hat{e}_i^\dag.
\end{align}
Focussing on $\hat{d}_j$, the frequency-domain expression is given by,
\begin{equation}
    -i\Omega \vec{d}(\Omega) = \vec{g}_d \hat{b}(\Omega)
    + \vec{g}_{d^\dag} \hat{b}^\dag(-\Omega)
    - iG \vec{d}(\Omega),
\end{equation}
where,
\begin{align}
    \vec{g}_d &= (ig_{d_1}, -ig_{d_1^\dag}^*, \dots, ig_{d_{n_d}}, -ig_{d_{n_d}^\dag}^*)^T, \\
    \vec{g}_{d^\dag} &= (ig_{d_j^\dag}, -ig_{d_j}^*, \dots, ig_{d_{n_d}^\dag}, -ig_{d_{n_d}}^*)^T,
\end{align}
and where,
\begin{equation}
    \vec{d}(\Omega) = (\hat{d}_1(\Omega),\dots,\hat{d}_{n_d}(\Omega); \hat{d}_1^\dag(-\Omega),\dots,\hat{d}^\dag_{n_d}(-\Omega))^T.
\end{equation}
and where in block form,
\begin{equation}
    G^{(d)} = \begin{bmatrix}G^{(d)}_1\\\vdots\\G^{(d)}_{n_d}\end{bmatrix} \in \mathbb{C}^{2n_d\times2n_d},
\end{equation}
where,
\begin{equation}
    G^{(d)}_j = \begin{bmatrix}
    g_{d_1 d_j^\dag} & \dots & g_{d_{n_d} d_j^\dag}; &
    g_{d_1 d_j} & \dots & g_{d_{n_d} d_j} \\
    g_{d_1 d_j^\dag}^* & \dots & g_{d_{n_d} d_j^\dag}^*; &
    g_{d_1 d_j}^* & \dots & g_{d_{n_d} d_j}^* \\
    \end{bmatrix},
\end{equation}
with $g_{d_j d_j} = g_{d_j d_j^\dag} = 0$. Solving for $\vec{d}(\Omega)$ gives,
\begin{align*}
    \vec{d}(\Omega) &= (-i\Omega I_{2{n_d}\times2{n_d}} + i G^{(d)})^{-1} \begin{bmatrix} \vec{g}_d & \vec{g}_{d^\dag}\end{bmatrix} \begin{bmatrix} \hat{b}(\Omega) \\ \hat{b}^\dag(-\Omega)\end{bmatrix} \\
    &\equiv M^{(d)} \begin{bmatrix} \hat{b}(\Omega) \\ \hat{b}^\dag(-\Omega)\end{bmatrix},
\end{align*}
where $I_{2{n_d}\times2{n_d}}$ is the $2{n_d}\times2{n_d}$ identity matrix, and $M^{(d)} \in \mathbb{C}^{2{n_d}\times 2}$.

The frequency domain expression for $\hat{b}$ is given by,
\begin{equation}
    -i\Omega \begin{bmatrix} \hat{b}(\Omega) \\ \hat{b}^\dag(-\Omega) \end{bmatrix}
    = \begin{bmatrix}i g_b & -i g_{b^\dag} \\ i g_{b^\dag}^* & -i g_b^* \end{bmatrix} \begin{bmatrix} \hat{a}(\Omega) \\ \hat{a}^\dag(-\Omega)
    \end{bmatrix}
    + i D^{(d)} \vec{d}(\Omega),
\end{equation}
where,
\begin{equation}
    D^{(d)} = \begin{bmatrix}
    g_{d_1},& \dots,& g_{d_{n_d}};& -g_{d_1^\dag},& \dots,& -g_{d_{n_d}^\dag} \\
    g_{d_1^\dag}^*,& \dots,& g_{d_{n_d}^\dag}^*;& -g_{d_1}^*,& \dots,& -g_{d_{n_d}}^*
    \end{bmatrix}.
\end{equation}
Solving for the $\hat{b}$ mode we get,
\begin{equation}
    \begin{bmatrix} \hat{b}(\Omega) \\ \hat{b}^\dag(-\Omega) \end{bmatrix} = T^{(b)}(\Omega) \begin{bmatrix} \hat{a}(\Omega) \\ \hat{a}^\dag(-\Omega)
    \end{bmatrix},
\end{equation}
where,
\begin{equation}
\label{eq:qcrb:closed-transfer-b}
    T^{(b)}(\Omega) \equiv (-i \Omega I_{2\times2} - i D^{(d)} M^{(d)})^{-1}
    \begin{bmatrix}i g_b & -i g_{b^\dag} \\ i g_{b^\dag}^* & -i g_b^* \end{bmatrix},
\end{equation}
and where $I_{2\times2}$ is the $2\times2$ identity matrix.

Similarly we have,
\begin{align*}
    \begin{bmatrix} \hat{c}(\Omega) \\ \hat{c}^\dag(-\Omega) \end{bmatrix} = T^{(c)}(\Omega) \begin{bmatrix} \hat{a}(\Omega) \\ \hat{a}^\dag(-\Omega)
    \end{bmatrix},
\end{align*}
where,
\begin{equation}
\label{eq:qcrb:closed-transfer-c}
    T^{(c)}(\Omega) \equiv (-i \Omega I_{2\times2} - i D^{(e)} M^{(e)})^{-1}
    \begin{bmatrix}i g_c & -i g_{c^\dag} \\ i g_{c^\dag}^* & -i g_c^* \end{bmatrix} 
\end{equation}
where,
\begin{equation}
    D^{(e)} = \begin{bmatrix}
    g_{e_1},& \dots,& g_{e_{n_e}};& -g_{e_1^\dag},& \dots,& -g_{e_{n_e}^\dag} \\
    g_{e_1^\dag}^*,& \dots,& g_{e_{n_e}^\dag}^*;& -g_{e_1}^*,& \dots,& -g_{e_{n_e}}^*
    \end{bmatrix},
\end{equation}
and where,
\begin{equation}
    M^{(e)} = (-i\Omega I_{2{n_e}\times2{n_e}} + i G^{(e)})^{-1},
\end{equation}
where in block form,
\begin{equation}
    G^{(e)} = \begin{bmatrix}G^{(e)}_1\\\vdots\\G^{(e)}_{n_e}\end{bmatrix} \in \mathbb{C}^{2n_e\times2n_e},
\end{equation}
where,
\begin{equation}
    G^{(e)}_j = \begin{bmatrix}
    g_{e_1 e_j^\dag} & \dots & g_{e_{n_e} e_j^\dag}; &
    g_{e_1 e_j} & \dots & g_{e_{n_e} e_j} \\
    g_{e_1 e_j^\dag}^* & \dots & g_{e_{n_e} e_j^\dag}^*; &
    g_{e_1 e_j}^* & \dots & g_{e_{n_e} e_j}^* \\
    \end{bmatrix}.
\end{equation}

The frequency domain expression for $\hat{a}$ is given by,
\begin{align*}
    -i\Omega \begin{bmatrix} \hat{a}(\Omega) \\ \hat{a}^\dag(-\Omega) \end{bmatrix} &= \begin{bmatrix} i g_b & -i g_{b^\dag} \\ i g_{b^\dag}^* & -i g_b^* \end{bmatrix}  \begin{bmatrix} \hat{b}(\Omega) \\ \hat{b}^\dag(-\Omega) \end{bmatrix} \\
    &+ \begin{bmatrix} i g_c & -i g_{c^\dag} \\ i g_{c^\dag}^* & -i g_c^* \end{bmatrix} \begin{bmatrix} \hat{c}(\Omega) \\ \hat{c}^\dag(-\Omega) \end{bmatrix} + \dots, \\
    &= T^{(a)}(\Omega) \begin{bmatrix} \hat{a}(\Omega) \\ \hat{a}^\dag(-\Omega) \end{bmatrix} + \dots
\end{align*}
where $\dots$ are the damping and input vacuum terms from Eq.~\eqref{eq:qcrb:langevin} and where,
\begin{align}
    T^{(a)}(\Omega) \equiv &\begin{bmatrix} i g_b & -i g_{b^\dag} \\ i g_{b^\dag}^* & -i g_b^* \end{bmatrix} T^{(b)}(\Omega)\nonumber\\ + &\begin{bmatrix} i g_c & -i g_{c^\dag} \\ i g_{c^\dag}^* & -i g_c^* \end{bmatrix} T^{(c)}(\Omega) = 0.
    \label{eq:qcrb:full-transfer-matrix}
\end{align}
Therefore to keep the input-output dynamics invariant, all elements of this matrix must be zero.

\end{document}